\documentclass[11pt,twoside]{article}

\usepackage{asp2006}
\usepackage{epsf}
\usepackage{psfig}
\usepackage{lscape}
\usepackage{graphicx}

\begin{document}
\title{Star Cluster Dissolution in Arp 284}  

\author{Bradley W. Peterson,\altaffilmark{1} Curtis Struck,\altaffilmark{1} Beverly J. Smith,\altaffilmark{2} and Mark Hancock\altaffilmark{2,3}}

\affil{\altaffilmark{1}Department of Physics and Astronomy, Iowa State University, Ames, IA 50011, USA} 
\affil{\altaffilmark{2}Department of Physics and Astronomy, East Tennessee State University, Johnson City, TN 37614, USA}
\affil{\altaffilmark{3}Institute of Geophysics and Planetary Physics, University of California, Riverside, Riverside, CA 92521, USA}

\begin{abstract}
We present results from a study of proto-globular cluster candidates in the interacting galaxy system Arp 284 (NGC 7714/5). Studies of the Antennae and M51 have suggested that the majority of these star clusters dissolve within 20~Myr due to mass loss. We obtain cluster colors using archival \emph{HST} data, and estimate ages and extinctions for over 150 clusters using evolutionary synthesis models. We find that clusters in NGC 7714 are generally less than 20~Myr old, while the data in the bridge is too limited to allow good estimates for individual clusters. We also examine {\mbox{H\,{\sc ii}}} region complexes with lower-resolution \emph{GALEX} and ground-based H$\alpha$ images. Several of these regions appear to be much older than the detected clusters within them, which may indicate the presence of an older, unresolved population of low mass or dispersed clusters.
\end{abstract}

\section{Introduction}
The study of extragalactic star clusters has benefited tremendously from the high resolution of the \emph{Hubble Space Telescope}. In interacting systems, where star clusters are particularly abundant, it is possible to study the demographics of cluster populations. Such studies have revealed an overabundance of young clusters in the Antennae \citep{fall05} and M51 \citep{bas05}, suggesting that clusters tend to dissolve due to mass loss after $\sim20$~Myr \citep{bas06,goo06}, an effect known as Òinfant mortality.Ó

In this study, we examine Arp 284 (NGC 7714/5), an interacting system at a distance of $38.6\pm2.7$ Mpc  ($1\arcsec$ = 190pc; NED). The system features several tidal environments that host ongoing star formation, including the starburst nucleus of NGC 7714, three tidal tails, and a two-component bridge. An old stellar ring in NGC 7714 and the post-starburst nucleus of NGC 7715 \citep{ber93} are no longer star-forming. More about the Arp 284 interaction may be found in the models of \citet{str03}.

We use high-resolution \emph{HST} data to study the star clusters and lower-resolution data from the \emph{Galaxy Evolution Explorer} (\emph{GALEX}) and ground-based H$\alpha$ to study the giant {\mbox{H\,{\sc ii}}} regions in which these clusters are found. A more detailed description of this project, including mid-IR data from \emph{Spitzer} and X-ray data from \emph{Chandra}, may be found in \citet{pet09}.

\begin{figure}[!ht]
\includegraphics[width=134mm]{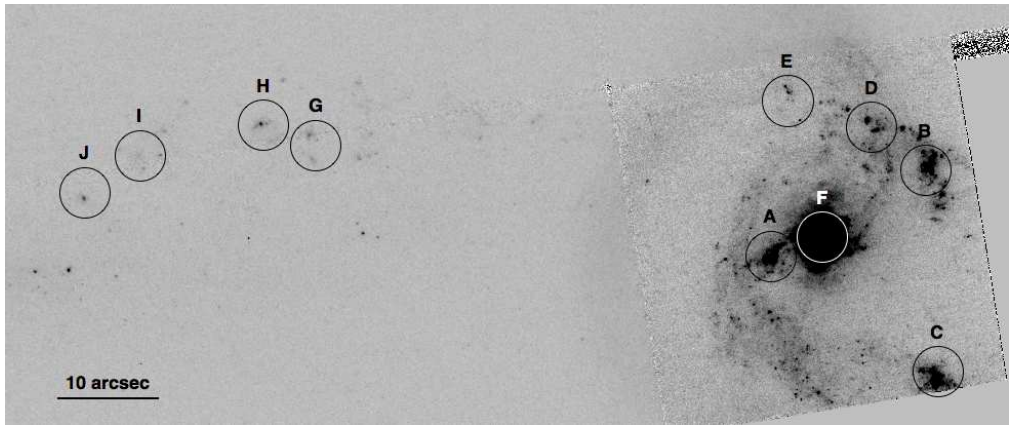}
\caption{\emph{HST} WFPC2 image of Arp 284 in F380W from the Hubble Legacy Archive. NGC 7714 is to the right, while NGC 7715 outside the image to the left. The giant {\mbox{H\,{\sc ii}}} regions, each of which contains numerous star clusters, are indicated. North is up and east to the left.
\label{fig:regions}}
\end{figure}

\section{Data and Analysis}
The \emph{HST} archives have images of Arp 284 taken with the Wide Field and Planetary Camera 2 (WFPC2) in four broadband filters. There are four images in F380W, two in F555W, one in F606W, and four in F814W. The observations do not provide uniform coverage of the system.

The ground-based H$\alpha$ data were obtained using the 1.8 m Perkins Telescope of the Ohio State University \citep{smi97}. The \emph{GALEX} data include a 4736 s exposure in the FUV (1516 \AA) and 520 s in the NUV (2267 \AA) bands.

Clusters were selected using the IRAF\footnote{IRAF is distributed by the National Optical Astronomy Observatories, which are operated by the Association of Universities for Research in Astronomy, Inc., under cooperative agreement with the National Science Foundation.} task \emph{daofind}, while aperture photometry was performed using the \emph{phot} task. The colors in the HST filters were then fit by $\chi^2$ minimization to models from the Starburst99 \citep{lei99,vaz05} evolutionary synthesis code, giving an estimate of the age and extinction for each cluster. All clusters was assumed to have formed in a single, instantaneous burst of star formation with a \citet{kro02} initial mass function. Models with metallicities of both $0.2Z_{\sun}$ and $0.4Z_{\sun}$ were fit. These abundances correspond to those measured in parts of the NGC 7714 disk by \citet{gon95}.

The giant {\mbox{H\,{\sc ii}}} regions (labeled A-I) were measured in the 4 \emph{HST} bands, 2 \emph{GALEX} bands, and H$\alpha$. The analysis was the same as for the \emph{HST} clusters, except that continuous star formation models were also fit. The apertures for these regions are shown in Figure~\ref{fig:regions}.

\section{Cluster Ages}
\begin{figure}[!ht]
\includegraphics[width=84mm]{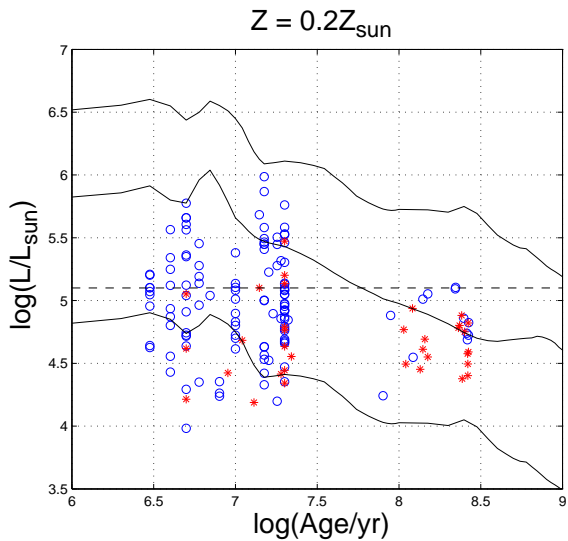}
\caption{F814W luminosity  vs. age for the $0.2Z_{\sun}$ models. Clusters in the bridge are shown as asterisks and those in the NGC 7714 disk are shown as circles. Nuclear clusters are not shown. The curves are Starburst99 evolutionary tracks for clusters with initial masses of $10^4$, $10^5$, and $5\times10^{5}M_{\sun}$. The horizontal dashed line marks the completeness limit.
\label{fig:lumVage}}
\end{figure}

The clusters in the tidal features of NGC 7714 are generally found to be young, with only 8 of 124 significantly older than 20 Myr. In contrast, the bridge appears to host an older cluster population, with nearly 50\% of the clusters older than 100 Myr. This is apparent in Figure~\ref{fig:lumVage}, which plots the luminosity of each cluster against its age for the $0.2Z_{\sun}$ model, with NGC 7714 clusters drawn as open circles and bridge clusters as asterisks. The difference between the populations is striking, but it should be noted that the ages in the bridge are more uncertain due to poorer coverage, higher photometric uncertainties, and the lack of metallicity data. New observations will be required to better determine the ages of the bridge clusters. If the older population is confirmed, it will be strong evidence that infant mortality has acted more weakly in the bridge than within the NGC 7714 disk and may suggest that environmental influences on early-stage cluster disruption are more important than presently believed.

Confirmation of infant mortality in Arp 284 is complicated by evolutionary fading. Clusters become dimmer as they age, so that only high mass clusters are detected at high ages. The completeness of a luminosity-limited sample is therefore also a function of the cluster ages. The curves in Figure~\ref{fig:lumVage} show the luminosity as a function of age for clusters with masses $10^4$, $10^5$, and $5\times10^{5}M_{\sun}$, while the dashed line is our completeness limit. Only the most massive clusters remain above our completeness limit after 100 Myr.

Due to the effects of evolutionary fading, our sample may be incomplete for older clusters, so the apparent effects of infant mortality in the system need confirmation.

\section{{\mbox{H\,{\sc ii}}} Region Ages}
The age estimates for the {\mbox{H\,{\sc ii}}} regions were determined using Starburst99 both by color fitting of the \emph{HST} and \emph{GALEX} data and by the H$\alpha$ equivalent widths. The EW(H$\alpha$) ages for the {\mbox{H\,{\sc ii}}} regions were generally quite young ($\la15$ Myr) and in good agreement with the median age of clusters found in the regions for those regions with more than a few clusters (A, B, D, and F), and were also in agreement with the ages determined spectroscopically by previous authors \citep{gar97, gon95, gon99}.

The color-based age estimates for regions B, D, and F were $\ga100$ Myr, significantly older than the median cluster ages, likely indicating a significant contribution from unresolved sources. These may be older clusters too faint to be detected or the remnants of a previous generation of clusters that has already dissolved.

\acknowledgements 
We thank the \emph{HST}, \emph{Spitzer}, \emph{GALEX}, and \emph{Chandra} teams for making this research possible. We acknowledge support from NASA Spitzer grant 1347980 and NASA Chandra grant AR90010B. BJS acknowledges support from NASA LTSA grant NAG5-13079. This research has made use of the NASA/IPAC Extragalactic Database (NED) which is operated by the Jet Propulsion Laboratory, California Institute of Technology, under contract with the National Aeronautics and Space Administration.

\end{document}